# Towards consensus on practical probabilistic evaluation metrics for clinical decision making in radiotherapy

Jenneke I. de Jong [a,b], Steven J.M. Habraken [c,b], Erik W. Korevaar [d], Mirko Unipan [e], Esther Kneepkens [e], Gloria Vilches Freixas [e], Yvonne L.B. Klaver [b,c], Michiel Kroesen [a,b], Alejandra Méndez Romero [a,b], Edwin Oldehinkel [d], Stephanie Peeters [e], Roel J.H.M. Steenbakkers [d], Karolien Verhoeven [e], Stefan Both [d], Liesbeth J. Boersma [e], Sebastiaan Breedveld [a], Mischa S. Hoogeman [a,b]

a. Radiotherapy, Erasmus MC Cancer Institute, Rotterdam, Netherlands.
b. Medical Physics & Informatics, HollandPTC, Delft, The Netherlands.
c. Radiation Oncology, Leiden University Medical Center, Leiden, The Netherlands.
d. Department of Radiation Oncology, University Medical Center Groningen, University of Groningen, Groningen, The Netherlands.
e. Maastro, Maastricht Radiation Oncology Institute, Maastricht, The Netherlands

---

**Abstract**

*Introduction*

The main aim of this paper is to identify clinically feasible probabilistic evaluation approaches for radiotherapy and to assess whether they can support clinical decision-making better than current robustness evaluation approaches. To this end, we calibrate probabilistic clinical goals and identify evaluation metrics with a team of medical physics experts and apply these to three patient cohorts.

*Methods*

Sixty intensity-modulated proton therapy plans from three Dutch proton centers were included: 20 neuro-oncological, 20 head-and-neck, and 20 breast cancer cases. Probabilistic evaluation was performed using polynomial chaos expansion with 10,000 simulated fractionated treatments per case. Volumetric and dosimetric CTV goals were calibrated at a 90% probability level. Selected probabilistic evaluation metrics were applied to representative patient cases and eight radiation oncologists participated in semi-structured interviews addressing preferred metrics, probability levels and clinical applicability.

*Results*

The calibrated volumetric goals were $D_{99.95\%,CTV}$, $D_{99.80\%,CTV}$ and $D_{99.54\%,CTV} = 0.95D_{pres}$ at a 90% probability level for the neuro-oncological, head-and-neck, and breast cohorts, respectively. The corresponding dosimetric goals were $D_{98\%,CTV} = 0.973D_{pres}$, $0.968D_{pres}$ and $0.967D_{pres}$ at a 90% probability level. All radiation oncologists considered the probabilistic evaluation metrics suitable for clinical decision-making, and most regarded them as an improvement over the currently used VWmin/max approach. Preferences for individual metrics and probability levels varied among radiation oncologists, centers and treatment sites.

*Conclusion*

Probabilistic evaluation metrics are expected to be clinically usable and an improvement over current methods by the interviewed radiation oncologists. Further consensus and clinical validation, in both proton and photon therapy, are needed before widespread clinical implementation.

---

## Introduction

In radiotherapy we aim for sufficient dose in the clinical target volume (CTV) while minimizing dose to surrounding organs at risk (OARs). This aim is translated into clinical goals that specify dosimetric or radiobiological requirements for the CTV and OARs.
In clinical practice, however, treatment delivery is subject to uncertainties, such as setup errors and anatomical changes, which can affect the delivered dose distribution. Therefore, treatment plans should be evaluated using evaluation metrics to determine whether the clinical goals remain satisfied in the presence of these uncertainties [1]. Evaluation metrics can consist of scalar quantities, such as DVH parameters, but may also represent higher-dimensional information, such as DVH curves or dose distributions.

Proton therapy is particularly sensitive to these uncertainties because of the finite proton range [2, 3]. Therefore, in intensity-modulated proton therapy (IMPT), these uncertainties are often addressed using robust optimization and evaluation techniques [4] rather than conventional PTV margins. Common practice involves simulating a set of error scenarios based on systematic setup and range errors. However, there is no international consensus on scenario-based evaluation, leading to variability across different centers [5]. According to a systematic review of the European Particle Therapy Network (EPTN) working group on treatment planning, common evaluation metrics for scenario-based evaluation include dose-volume histogram (DVH) bands and the range of DVH values within the error scenario set [6]. In addition, voxel-wise minimum (VWmin) and maximum (VWmax) dose distributions and the associated DVH values are often presented [6]. In the Netherlands, scenario-based evaluation has been standardized among the three proton centers, involving the aggregated VWmin and VWmax dose distributions derived from 28 error scenarios [7]. The error scenarios are selected on the surface of a sphere with a radius equal to the treatment-site-specific clinical margin.

Although these methods provide insight into plan robustness, they do not provide insight into the dosimetric outcomes at various well-defined probability levels. Probabilistic evaluation [8, 9], which explicitly incorporates the probability distributions of setup and range errors and models both systematic and random errors, enables such assessment. While probabilistic evaluation typically requires more dose calculations, recent methods have enabled its implementation with computation times comparable to current robustness evaluation methods [8, 10]. Previous research has shown that probabilistic evaluation could improve the trade-off between OAR sparing and target coverage [11] and reduce the interpatient variation in probabilistic outcomes [12]. In the ESTRO-EPTN Delphi consensus paper about robustness evaluation in proton therapy, the implementation of probabilistic evaluation received the highest priority score for clinical integration [1].

Several evaluation metrics for probabilistic evaluation have been suggested, such as probability distributions of DVH metrics [9, 13, 14], DVH values belonging to specific probabilities [11, 12, 14, 15], DVH bands between selected probabilities [12, 14, 16-19], dose-probability histograms [20], probability distributions of normal tissue complication probability (NTCP) values [20] and probability maps belonging to a specific dose [18].

Although both the clinical benefits and technical feasibility of probabilistic evaluation are promising, its integration into routine clinical workflows remains limited to date. This is, at least partly, due to:

1. **Lack of standardized probabilistic clinical goals for use in clinical practice.** Calibrating a probabilistic CTV goal based on current clinical goals is not straightforward and it is important to understand the effect of different calibration and probability level choices. The need for standardization is further emphasized by the considerable heterogeneity in probability levels currently used in probabilistic planning studies [21].
2. **Lack of standardized probabilistic evaluation metric for use in clinical practice.** Probabilistic evaluation is based on a large number of dose distributions and intuitive metrics are needed to summarize this information in a way that supports clinical decision-making.

The main aim of this paper is to calibrate clinically feasible probabilistic clinical goals and identify evaluation metrics for radiotherapy and to assess whether they can support clinical decision-making better than current robustness evaluation methods. To this end, we identify probabilistic clinical goals and evaluation metrics with a team of medical physics experts and apply these to three patient cohorts from different centers and treatment sites. Subsequently, we assess their clinical usability through semi-structured interviews with radiation oncologists in a multicenter setting.

## Materials and methods

### *Clinical treatment plans*

A total of 60 treatment plans from patients treated at the three Dutch proton centers were included. All plans were created in RayStation using the clinical beam model of the individual centers and the Monte Carlo (MC) dose engine with a MC uncertainty of 1%.

For center 1, 20 treatment plans from neuro-oncological patients treated in 2023 were included. The prescribed dose ($D_{pres}$) was 50.4 Gy RBE (16/20) or 59.4 Gy RBE (4/20), delivered in 1.8 Gy RBE fractions using a Varian ProBeam system. All clinical plans were generated using a single-field optimization IMPT (SFO-IMPT) approach using patient-specific non-coplanar arrangements of two or three beams.

For center 2, 20 treatment plans from head-and-neck cancer patients treated in 2021 were included. $D_{pres}$ was 70 Gy RBE to the primary CTV (macroscopic tumor) and 54.25 Gy RBE to the elective CTV (lymph nodes), delivered in 35 fractions using the IBA Proteus PLUS system. Treatment plans were made using beam angle configurations of five to six beams, primarily with anterior and posterior oblique coplanar directions, and subsequent manual fine-tuning of the beam angles per patient. All clinical plans were generated using a multi-field optimization IMPT (MFO-IMPT) approach. Posterior fields were used without a range shifter.

For center 3, 20 treatment plans from breast cancer patients treated in 2023-2025 were included. $D_{pres}$ was 40.05 Gy RBE delivered in 2.67 Gy RBE fractions to the primary CTV (8 breast and 12 thoracic wall), internal mammary nodes (16 patients) and the level 1-4 (18 patients) or level 3-4 (2 patients) axillary lymph nodes using the Mevion S250i system. All clinical plans were generated using a MFO-IMPT with patient-specific arrangements of two to four beams.

*Clinical robust optimization and evaluation methods*

In all three centers, robust CTV dose under setup and range errors was aimed for through minimax optimization [22] with 21 error scenarios that were formed by a combination of a ±3 mm (neuro-oncological and head-and-neck) or ±5 mm (breast) setup error in each of the three principal directions and a stopping-power prediction (range) error of −3%, 0%, or +3% . Robust optimization was applied to the CTV objectives and critical OAR constraints, guiding the optimizer to achieve the desired outcomes across all error scenarios. The remaining OAR objectives were only optimized in the nominal scenario.

The clinical plans were evaluated based on 28 error scenarios [7]. These error scenarios included 14 setup error combinations with directions determined by the 6 faces and 8 vertices of a cube, all normalized to 3 mm (neuro-oncological and head-and-neck) or 5 mm (breast). The radius of the sphere defining the setup error magnitude corresponds to the treatment-site-specific van Herk margin used in photon therapy.

The setup errors were combined with range errors of −3% or 3%. CTV dose was evaluated based on the VWmin dose distribution of the 28 error scenarios. The clinical goals for the CTV were VWmin $D_{98\%,CTV}$=0.95$D_{pres}$, VWmin $D_{98\%,CTV}$=0.94$D_{pres}$ and VWmin $D_{98\%,CTV}$=0.93$D_{pres}$ in center 1, 2 and 3, respectively, based on the center-specific calibrations described in [7]. Serial OAR doses were evaluated using the VWmax dose distribution of the 28 error scenarios, whereas non-serial OAR doses were evaluated in the nominal dose distribution.

*Probabilistic evaluation*

To perform a fast probabilistic evaluation for a treatment plan, a Polynomial chaos expansion (PCE) model was created and 10,000 fractionated treatments were simulated. PCE is a plan-specific model of the dose engine based on polynomial functions, validated for radiotherapy applications in [8]. For every treatment, a systematic setup error and a range error were randomly sampled from their Gaussian distribution and applied to all fractions within a treatment course. Additionally, for each fraction a randomly sampled random setup error was applied.  For each center, the mean and standard deviation (SD) of the errors that were used in the PCE models are presented in Table 1.

The means and SD values for the setup uncertainties were chosen to align with the clinical margin $M$ used for robust optimization and evaluation scenario selection. The values were derived based on Van Herk's margin recipe: $M = 2.5\Sigma + 0.7\sigma$ [23], with $\Sigma$ the SD of the systematic error and $\sigma$ the SD of the random error. A relative contribution of systematic and random errors was assumed, with the systematic error slightly smaller than the random error, consistent with the approach described by Rojo-Santiago, J., et al. [9], leading to SD values close to values measured in former measurements in each center.

The SD of the range uncertainty depended on the treatment site. Range uncertainty reflects uncertainties in CT-based stopping-power prediction, including the center-specific CT-to-SPR calibration [24]. For the neuro-oncological patients, the CT-value to proton stopping power prediction (SPP) calibration of the single-energy CT (SECT) acquisition and reconstruction protocol was based on measurements with a Gammex 467 phantom. In neuro-oncology treatments with SECT this leads to a systematic underestimation of proton SPP of 1.2% ± 0.7% (1 SD) [25], i.e., a range undershoot, as compared to Dual-energy CT (DECT). We assumed an additional uncorrelated SPP error in DECT of 0.7% [26] and arrived at a total systematic SPP error of 1.2% ± 1.0% (1 SD). For the head-and-neck and breast patients, a SD of 1.5% for ρ was chosen, consistent with values reported in literature [27].

**Table 1.** *Choice of means and standard deviations for the errors used in probabilistic evaluation.*

| Center | Treatment site | Clinical Margin ($M$) | SD systematic setup error ($\Sigma$) | SD Random setup error ($\sigma$) | Mean ± SD Systematic range error ($\rho$) |
|---|---|---|---|---|---|
| Center 1 | Neuro-oncological | 3 mm | 0.92 mm | 1.00 mm | 1.2 ± 1.0% |
| Center 2 | Head-and-neck | 3 mm | 0.92 mm | 1.00 mm | 0.0 ± 1.5% |
| Center 3 | Breast | 5 mm | 1.50 mm | 1.90 mm | 0.0 ± 1.5% |

*Calibration of probabilistic clinical goals*

A team of medical physicists from the three participating proton centers proposed methods for probabilistic clinical goal calibration and selected those considered clinically feasible. The selected methods were applied to patient cohorts and discussed in interviews with radiation oncologists.

The van Herk criterion, which states that the point minimum dose to the CTV should be $0.95D_{pres}$ with a 90% probability level [23], served as the reference framework for calibration. Because the absolute minimum dose is sensitive to calculation uncertainty and single-voxel effects, $D_{98\%,CTV}$ is commonly used in clinical practice as a practical surrogate for the minimum CTV dose [28]. For each calibration approach, two of the three parameters (dose, volume, and probability) were fixed based on the van Herk criterion and this clinical surrogate, while the third parameter was calibrated from 20 clinical treatment plans per treatment site (Fig. 1).

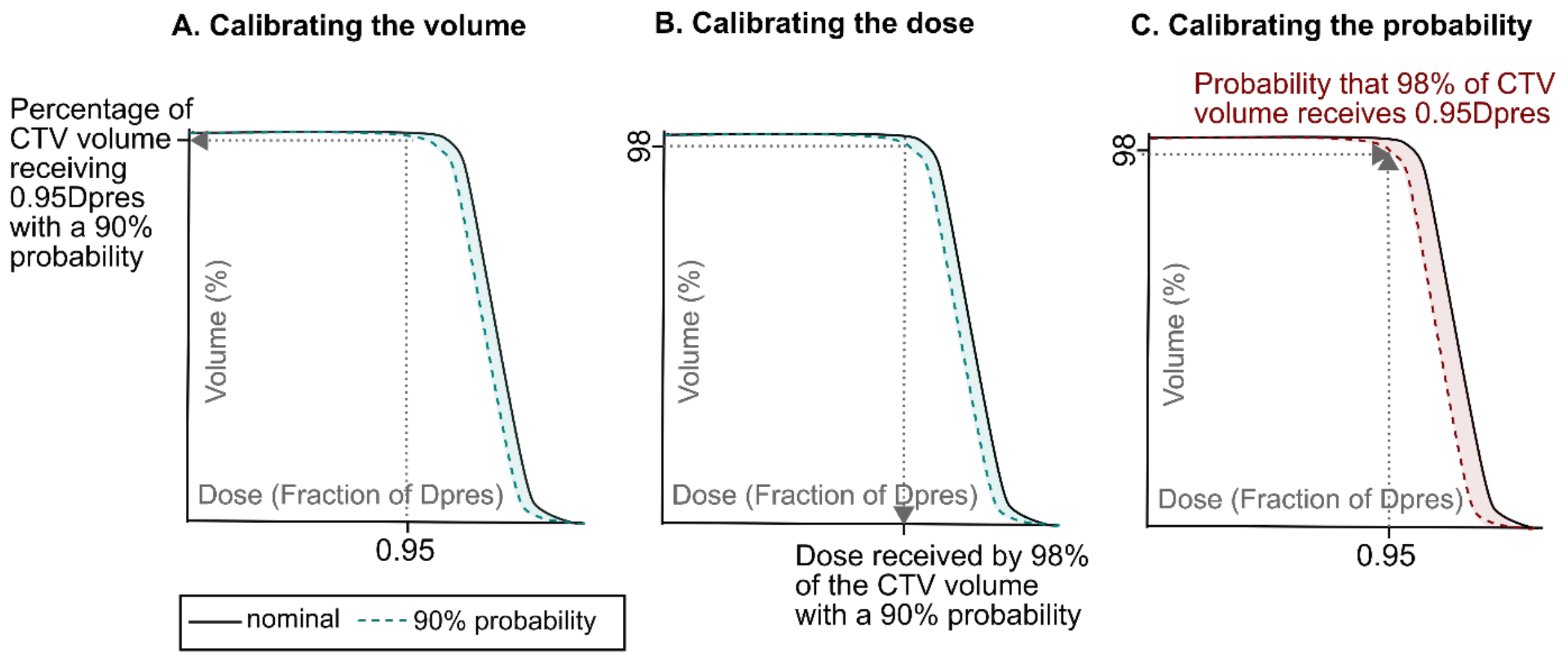

**Fig. 1.** *Overview of the different options for probabilistic CTV goal calibration. In volumetric calibration (A), the dose level and probability are chosen and the corresponding CTV volume is determined by calibration. In dosimetric calibration (B), the CTV volume and probability are chosen and the corresponding dose level is determined by calibration. In probability calibration (C), the dose level and CTV volume are chosen and the corresponding probability level is determined by calibration. Solid lines represent the nominal DVH, and dashed lines represent the DVH corresponding to the selected probability level.*

*Identification of evaluation metrics*

A team of medical physicists from the three participating proton centers proposed probabilistic evaluation metrics and selected those considered clinically feasible.

Probabilistic evaluation of DVH parameters can range from full probability distributions to summary metrics. A comprehensive representation can be obtained by presenting the full probability distribution of each DVH parameter (Fig. 2A). In addition, summary metrics can be derived, such as the dose corresponding to a given DVH parameter at a specified probability level (Fig. 2B) or the probability of achieving a clinical goal (Fig. 2C).

DVH curves can also be extended to a probabilistic framework by incorporating uncertainty bands that represent selected probability levels (Fig. 2D). Alternatively, modified DVH-type plots can be constructed by replacing either the dose or volume axis with probability, thereby visualizing the probability of achieving a given dose or volume threshold (Fig. 2E and 2F).

At the spatial level, probabilistic 3D distributions can be generated in different ways. One approach is to fix a probability level and assign to each voxel the corresponding dose value (i.e. the dose achieved with that probability) (Fig. 2G). This results in an aggregated dose distribution, similar to the VWmin/max dose distribution used in clinical practice, and does not necessarily represent a physically realizable treatment scenario. Alternatively, probability maps can be constructed that represent the probability of achieving a specified dose level in each voxel (Fig. 2H).

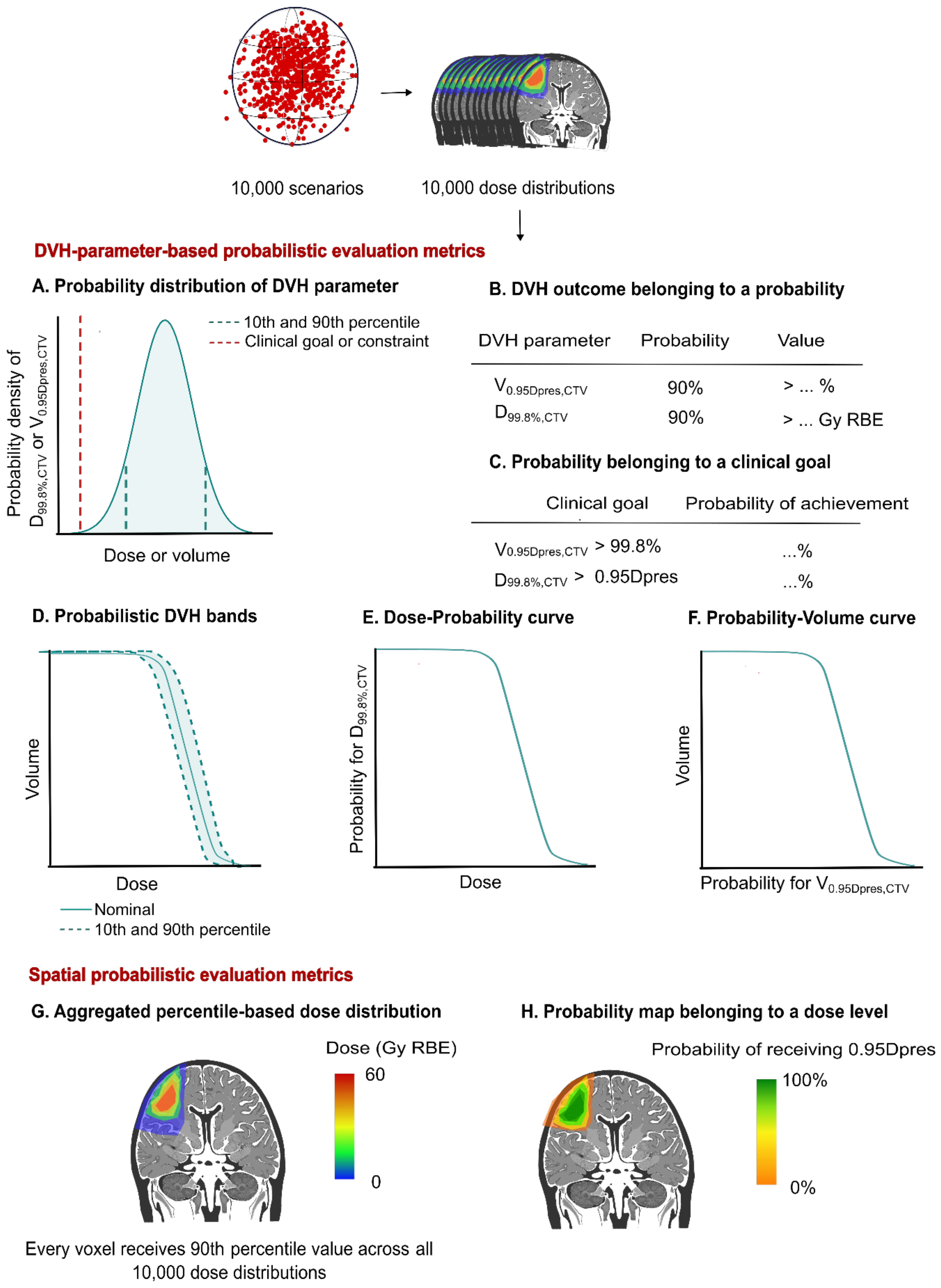

**Fig. 2.** *Overview of the different options for probabilistic evaluation metrics. DVH-parameter–based metrics include (A) the full probability distribution of a DVH parameter, (B) the DVH value*

*corresponding to a selected probability level, and (C) the probability of achieving a specified clinical goal. DVH-based representations include (D) probabilistic DVH bands and (E–F) dose- and volume-probability representations. Note that E and F represent the same information as A in a different visualization format. Spatial representations include (G) the dose corresponding to a selected probability level in each voxel and (H) the voxel-wise probability of achieving a specified dose level. Examples are shown for the CTV but could also be applied to OARs. In that case, the probability level would change and the direction of the signs (> versus <) would be reversed where applicable.*

*Application to patient cases*

Before calibration of the probabilistic goals, the 60 clinical plans in the three cohorts were scaled to precisely meet the current clinical VWmin clinical goal for the CTV, ensuring that the probabilistic calibration was directly based on these goals and that all clinical variation was removed.

For the probabilistic CTV goal, calibration was performed at a fixed probability level of 90% using both the volumetric (Fig 1A) and dosimetric (Fig 1B) approach. For each patient, probability distributions of $V_{95\%,CTV}$ and $D_{98\%,CTV}$ were generated using PCE. The 10$^{th}$ percentile of each distribution was extracted, representing the value achieved with a 90% probability level. Population-level probabilistic goals were then defined as the median of these values across all treatment plans. The probabilistic OAR constraints are based on meeting the constraints used in clinical practice with a 95% probability level. The distributions and individual patient values were visualized using boxplots. Differences among the three cohorts were assessed using two-sided Welch's independent-samples t-tests with a significance level of $\alpha$=0.05.

For the evaluation metrics, a subset of the identified probabilistic evaluation metrics (Fig 2A-2C and 2G) was applied to representative patient cases from each center. Fig. 2B, 2C, and 2G were chosen because they closely correspond to the metrics currently used in the clinic based on 28 scenarios. Specifically, Fig. 2B replaces VWmin and VWmax values with values belonging to a specified probability level, Fig. 2C replaces the number of scenarios meeting a clinical goal with the probability of achieving that goal, and Fig. 2G replaces VWmin and VWmax dose distributions with percentile-based voxel-wise dose distributions. Fig. 2A, while not analogous to existing metrics, was thought to provide a logical way to represent the full probability distribution of each DVH parameter and was therefore included.

For Fig. 2B, the 10$^{th}$ percentile values of CTV DVH parameters and the 95$^{th}$ percentile values of OAR DVH parameters were reported. For Fig. 2G, the dose distributions were created for the 10$^{th}$ and 95$^{th}$ percentile dose values for analysis of probabilistic target coverage and OAR doses, respectively.

*Interviews with radiation oncologists*

Semi-structured interviews with eight radiation oncologists, three each from center 1 and 3 and two from center 2, were conducted. The full set of interview questions is provided in Table 2. The radiation oncologists were asked to express their preference for volumetric (Fig. 1A) versus dosimetric (Fig. 1B) calibration and to provide feedback on the selected probability levels for the clinical goals and OAR constraints. In addition, the radiation oncologists reviewed the proposed probabilistic evaluation metrics (Fig. 2A-2C and 2G) and were asked to indicate their preferred method, as well as whether these approaches could support clinical decision making and replace the conventional VWmin/max evaluation method.

## Results

*Calibration of probabilistic clinical goals*

For the 20 neuro-oncological patients, the median $V_{0.95Dpres,CTV}$ was 99.95% (range, 99.84%–100.00%), corresponding to a volumetrically calibrated clinical goal of $D_{99.95\%,CTV} = 0.95D_{pres}$ with 90% probability. The median $D_{98\%,CTV}$ was $0.973D_{pres}$ (range, $0.967D_{pres}$–$0.982D_{pres}$), corresponding to a dosimetrically calibrated goal of $D_{98\%,CTV} = 0.973D_{pres}$ with 90% probability.

For the 20 head-and-neck patients, the median $V_{0.95Dpres,CTV}$ was 99.80% (range, 99.69%–99.92%), resulting in a volumetrically calibrated goal of $D_{99.80\%,CTV} = 0.95Dpres$ with 90% probability. The median $D_{98\%,CTV}$ was $0.968D_{pres}$ (range, $0.961D_{pres}$–$0.973D_{pres}$), resulting in a dosimetrically calibrated goal of $D_{98\%,CTV} = 0.968D_{pres}$ with 90% probability.

For the 20 breast patients, the median $V_{0.95Dpres,CTV}$ was 99.54% (range, 98.91%–99.81%), resulting in a volumetrically calibrated goal of $D_{99.54\%,CTV} = 0.95D_{pres}$ with 90% probability. The median $D_{98\%,CTV}$ was $0.967D_{pres}$ (range, $0.954D_{pres}$–$0.979D_{pres}$), resulting in a dosimetrically calibrated goal of $D_{98\%,CTV} = 0.967D_{pres}$ with 90% probability. The values for all patients are presented in Figure 3.

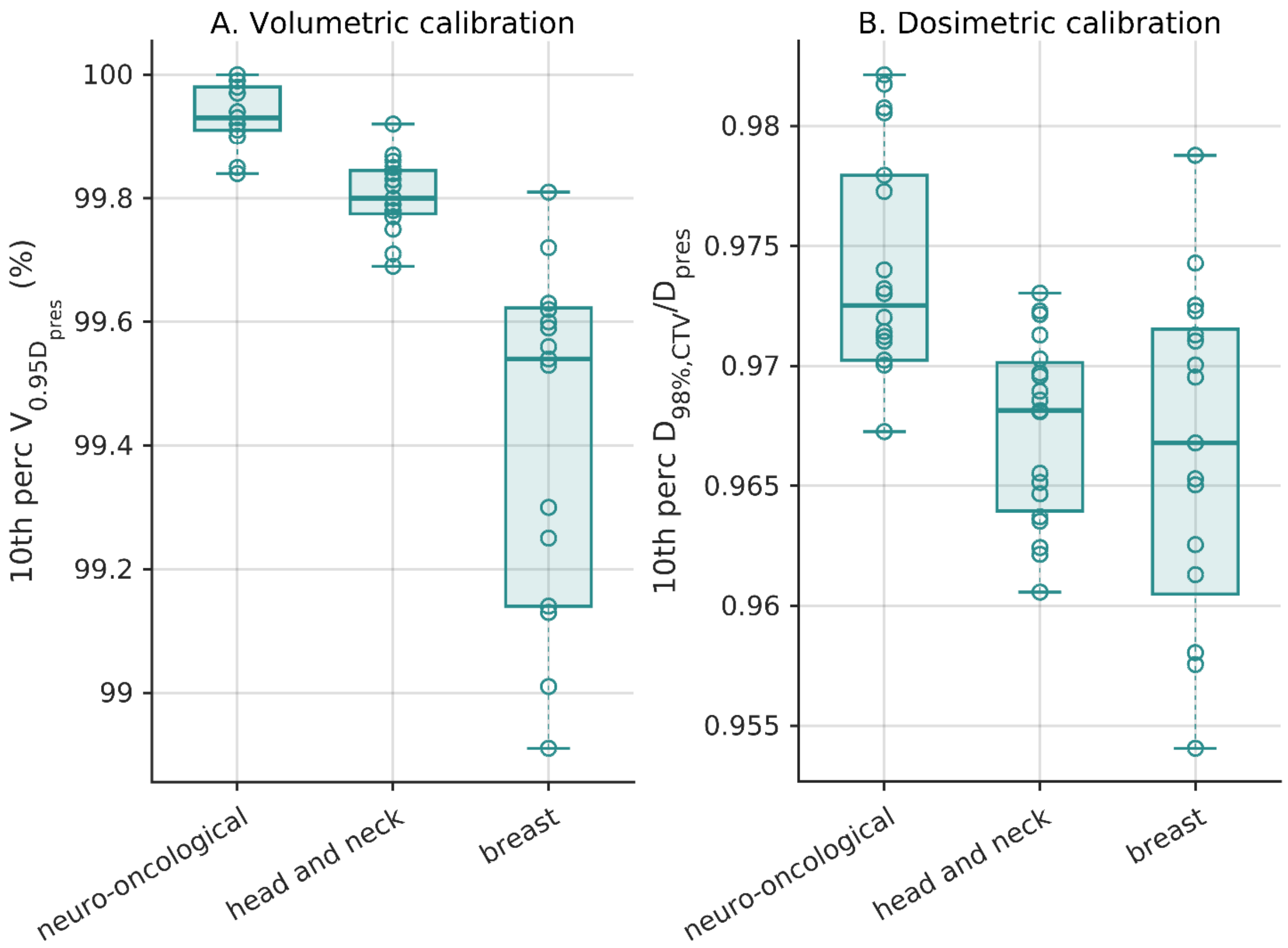

**Fig. 3.** *Interpatient variation in the $V_{0.95Dpres,CTV}$ and $D_{98\%,CTV}$ met with a 90% probability across the three patient cohorts. Individual points represent patients, boxes indicate the interquartile range, and*

*the thick horizontal lines indicate the median values that were used to derive the population-level probabilistic CTV goals.*

Welch's t-tests showed statistically significant differences for all group comparisons at the 0.05 significance level, except for the dosimetric outcomes of the head-and-neck and breast cohorts ($p=0.611$).

*Application to patient cases*

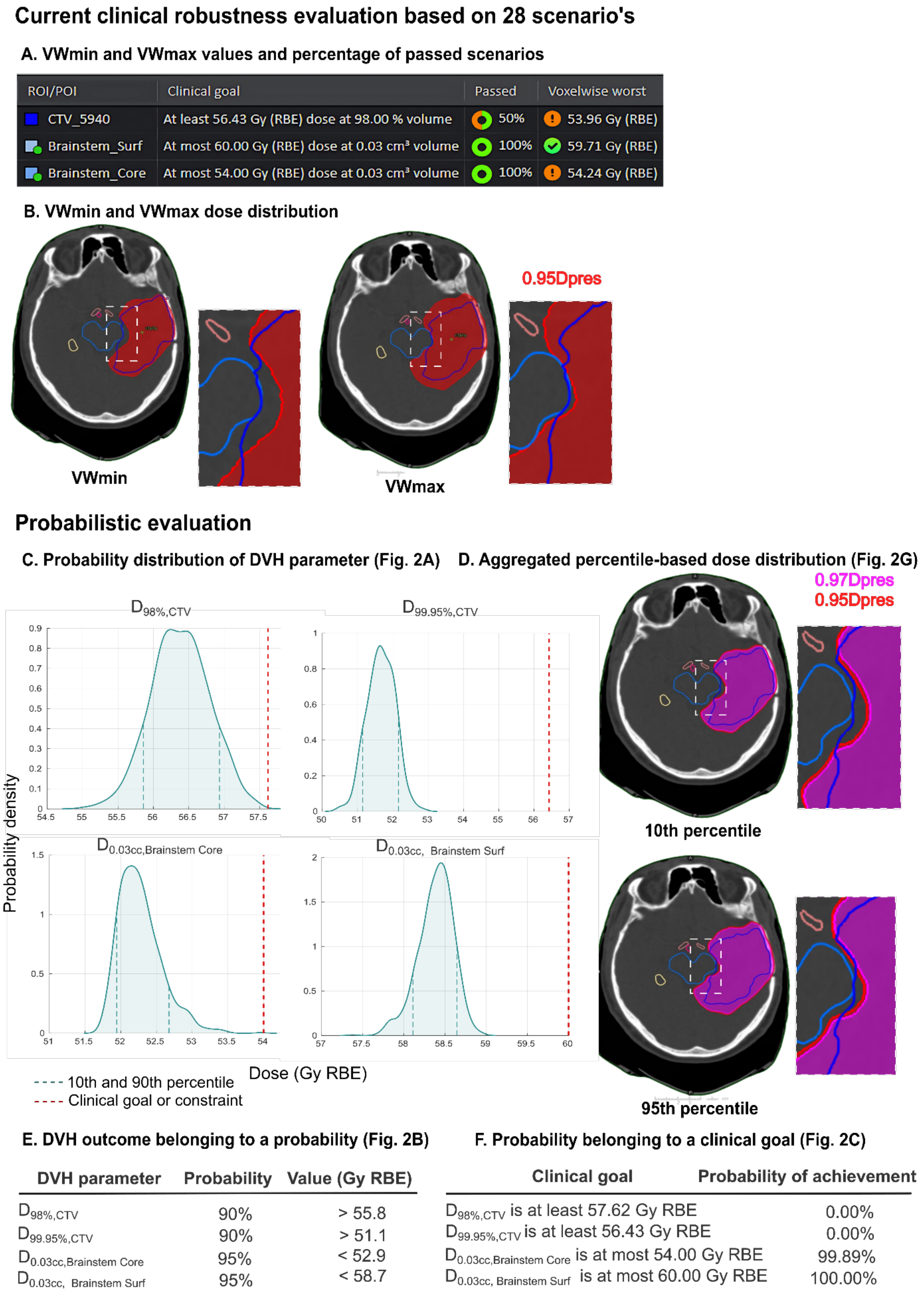

**Fig. 4.** *Comparison of conventional scenario-based robustness evaluation and selected probabilistic evaluation metrics for a representative neuro-oncological case used in the interviews. (A) Current clinical evaluation showing the VWmin/VWmax values and the percentage of the 28 error scenarios satisfying each clinical goal. (B) VWmin and VWmax dose distributions, including enlarged views of the target–OAR interface. (C) Probability distributions of selected CTV and brainstem DVH parameters. (D) Voxel-wise percentile dose distributions corresponding to the 10th and 95th percentiles, used for evaluation of target coverage and OAR dose, respectively. Enlarged views illustrate the corresponding isodose distributions near the target–OAR interface. (E) DVH parameter values corresponding to specified probability levels. (F) Probabilities of achieving the specified clinical goals.*

Figure 4 shows the probabilistic evaluation metrics applied to the neuro-oncological case used in the radiation oncologist interviews. Similar evaluations were performed for representative patient cases from the other treatment sites and centers and were also discussed during the interviews.

*Interviews with radiation oncologists*

Table 2 summarizes the interview results. Superscripts in the following text are used to refer to the corresponding entries in the table. No radiation oncologist expressed a strong preference for either volumetric or dosimetric calibration (Q1).[1] Four of five neuro-oncology and head-and-neck radiation oncologists slightly preferred volumetric calibration (Fig. 1A), primarily because maintaining 95% of the prescribed dose more closely resembles photon planning practice.[2] However, one head-and-neck radiation oncologist and all breast radiation oncologists noted that underdosing the worst 2% of the CTV volume is of limited clinical relevance.[3] One breast radiation oncologist nevertheless preferred volumetric calibration because it avoids detailed evaluation of the lowest-dose region of the CTV.[4]

Three of eight radiation oncologists supported the use of sequential CTV goals (Q2), as this reflects previous clinical practice[5] and indicates what target coverage remains achievable when the primary goal is not met.[6]

Preferences regarding the probability level for the probabilistic CTV goal (Q3) were evenly divided. Four radiation oncologists preferred 90%, citing consistency with the van Herk margin concept and available clinical data.[8] The remaining radiation oncologists favored a higher probability, with two mentioning that steeper dose gradients in proton therapy, when there is a trade-off, might cause substantial underdosage in the worst 10% of scenarios.[7]

The three radiation oncologists from center 1 preferred calibration against photon plans (Q4), as their current proton-planning approach was considered conservative.[9] In contrast, four of five radiation oncologists from centers 2 and 3 expected calibration based on photon and proton plans to yield equivalent results and recommended evaluating both to confirm consistency.[10] One radiation oncologist proposed calibration based on tumor control probability, arguing that matching photon and proton doses might result in unnecessarily high proton doses because of the greater inherent robustness of photon plans.[11]

Five of eight radiation oncologists indicated that the probability threshold for OAR constraints (Q5) should depend on the organ. Two considered 95% sufficient for critical OARs, such as the brachial plexus or brainstem, while accepting lower thresholds for non-critical OARs. Conversely, three considered 95% appropriate for non-critical OARs and preferred thresholds of 98–99% for critical OARs.[12]

Regarding probabilistic evaluation metrics (Q6), opinions on the clinical value of the full probability distribution (Fig. 2A) were divided. Four radiation oncologists found it difficult to interpret or insufficiently specific for clinical decision making,[13] whereas four considered it useful for evaluating selected trade-offs.[14] All radiation oncologists supported the use of a summary metric. The DVH value corresponding to a specified probability (Fig. 2B) and the probability of meeting a clinical goal (Fig. 2C) were similarly preferred, with no clear overall preference.
Two radiation oncologists would choose to examine multiple percentiles for the CTV (Q7) to understand the impact of increasing or decreasing the probability of target coverage[15], while three radiation oncologists preferred examining multiple percentiles for the OARs to gain more insight into the tails of the distributions.[16]
The radiation oncologists unanimously agreed that the probabilistic evaluation metrics (Q8) are suitable for clinical decision making. Half of the radiation oncologists noted that these metrics provide more information or better insight compared to the VWmin/max method currently in use[17]. Two radiation oncologists observed that the new method is less conservative than the current approach[18].

Six of eight radiation oncologists also supported the clinical use of probabilistic dose distributions (Q9), while two were uncertain. One of the latter acknowledged their potential but found them difficult to interpret in practice and preferred the simplicity of margin-based methods.[19]

**Table 2.** *The table summarizes all results of the radiation oncologist interviews. Superscripts in the table refer to the corresponding descriptions in the text of this section.*

| | ***Radiation oncologist 1 Neuro Center 1*** | ***Radiation oncologist 2 Neuro Center 1*** | ***Radiation oncologist 3 HNC Center 1*** | ***Radiation oncologist 4 HNC Center 2*** | ***Radiation oncologist 5 HNC Center 2*** | ***Radiation oncologist 6 Breast Center 3*** | ***Radiation oncologist 7 Breast Center 3*** | ***Radiation oncologist 8 Breast Center 3*** |
|---|---|---|---|---|---|---|---|---|
| **Probabilistic CTV goal and OAR constraint calibration** | | | | | | | | |
| *Q1: Do you prefer the probabilistic CTV goal that follows from volumetric (Fig. 1A) or from dosimetric calibration (Fig. 1B)?* [1] | volumetric[2] | volumetric[2] | volumetric | volumetric | dosimetric[3] | volumetric[3,4] | no preference [3] | no preference[3] |
| *Q2: Would you prefer having multiple clinical goals for the CTV?* | unsure | yes[5] | yes | unsure | yes[6] | - | - | - |
| *Q3: What probability should be in the CTV goal?* | 95/98 | 98[7] | 95 | 90[8] | 90[8] | 90/95[7] | 90[8] | 90[8] |
| *Q4: Should we calibrate based on current clinical practice or on photon planning?* | photon plans [9] | photon plans [9] | photon plans | both[10] | both[10] | based on TCP models[11] | both[10] | both[10] |
| *Q5: What probability should be in the OAR constraints?* | 95 or lower [12] | 95/98[12] | 95 | 90 | 95/98[12] | 95/99[12] | 95 or lower[12] | 95 |
| **Probabilistic evaluation metrics (Fig. 2A/B/C)** | | | | | | | | |
| *Q6: What metric or combination of metrics for evaluation of the clinical goals would you prefer?* | *Fig. 2B/ Fig. 2C* [13] | *Fig. 2B/ Fig. 2C* [13] | *Fig. 2A/ Fig. 2C*[14] | *Fig. 2B* [13] | *Fig. 2B/ Fig. 2C* [13] | *Fig. 2A/ Fig. 2B*[14] | *Fig. 2A/ Fig. 2C* [14] | *Fig. 2A/ Fig. 2B*[14] |
| *Q7: What percentiles would be interesting to gain insight into for Fig. 2B?* | *95/98*[15] | *95/98/100*[16] | *-* | *90* | *95/98*[16] | *95/99*[16] | *95* | *90/95*[15] |
| *Q8: Can the probabilistic evaluation metrics be used for clinical decision making and replace the VWmin/max method?* | *yes* | *yes*[17] | *yes* | *yes* | *yes*[17] | *yes*[17] | *yes*[18] | *yes*[17, 18] |
| **Probabilistic evaluation metrics (Fig. 2G)** | | | | | | | | |
| *Q9: Can the probabilistic dose distribution be used for clinical decision making and replace the VWmin/max method?* | *yes* | *unsure/yes*[19] | *unsure* | *yes* | *yes* | *yes* | *yes* | *yes* |

## Discussion

In this paper, we assessed the clinical feasibility of probabilistic clinical goals and evaluation metrics through interviews with radiation oncologists. To support this assessment, a team of medical physics experts first identified and further developed clinically feasible probabilistic clinical goals and evaluation metrics, which were subsequently applied to patient cases across different treatment sites and centers. To the best of our knowledge, although most of the evaluated metrics have been described previously [9, 12-20], this is the first multi-institutional study to investigate radiation oncologists' views on the practical application of probabilistic evaluation. Our main finding is that the probabilistic clinical goals and evaluation metrics are perceived as clinically usable according to all eight radiation oncologists who were interviewed. The majority of radiation oncologists expected them to be an improvement over the current VWmin/max method, which could be explained by the need for additional information in IMPT, where complex trade-offs between target coverage and OAR sparing frequently need to be made. In contrast, preferences regarding the specific probability levels defining clinical acceptability showed substantial variation.

Previous radiotherapy consensus studies have similarly used clinician interviews and expert discussions to identify relevant issues and formulate candidate recommendations before these were subjected to formal (Delphi) consensus procedures [29-31]. Our study may therefore be considered a first step toward consensus on probabilistic evaluation, by identifying clinically feasible approaches and areas in which radiation oncologists' preferences differ.

Regarding the probabilistic evaluation metrics, there was substantial inter-radiation oncologist variation in preferences regarding the evaluation metrics. This variation is not expected to pose practical limitations, as Fig. 2B and Fig. 2C can be easily combined to accommodate most clinical cases, with Fig. 2A available when additional information is required.
Regarding the probabilistic CTV goal, radiation oncologists preferred volumetric calibration over dosimetric calibration, although these preferences were not strong, with an emphasis on maintaining consistency with photon planning. Radiation oncologists did mention that they expected that the clinical impact of either approach would be minimal, suggesting that the method favored by the experts could be selected without further comparative evaluation.
Because volumetric calibration outcomes differed significantly between all cohorts (Fig. 3A), treatment-site-specific volumetric goals may be more appropriate than a single common goal, although the clinical relevance of these differences should also be considered. It should be noted that these differences cannot be attributed solely to treatment site, as the cohorts also differed in other aspects, including center, machine and planning technique.
Almost all radiation oncologists agreed that photon plans should be included in the calibration process. This preference may partly reflect the Dutch clinical practice, where most proton therapy indications are determined using model-based comparisons with photon therapy [32]. TCP-based calibration may be an improvement in the future once clinically validated TCP models for IMPT are available for all treatment sites.

Substantial variation was observed in preferences regarding the probabilistic CTV goal and OAR constraint probability levels, reflecting differences between centers, individual radiation oncologists and treatment sites. Several radiation oncologists indicated that broader consensus is desired and reported difficulty in providing definitive answers on an individual basis. This uncertainty likely stems from limited evidence linking specific probability levels to clinical outcomes. Future work, based on probabilistic evaluation, could include a planning study examining the worst 10 percent of simulated treatment outcomes when using a 90 percent probability level for the CTV goal, specifically whether these scenarios still receive doses near the clinical goal or result in significant underdosing, and whether this differs for IMPT compared to photon planning. Additionally, the effect of increasing the CTV coverage probability above 90 percent on OAR dose should be assessed for both modalities. Ultimately, a standardized set of probability levels may be established for the CTV and individual OARs, while retaining some flexibility in cases involving trade-offs.

Eventually, probability distributions of TCP and NTCP could be directly presented in the TPS to enhance objectivity and support data-driven decision making. While NTCP models exist for the relevant treatment sites [33-37], it is important to remain aware of extra uncertainties introduced by modeling. Both the expected TCP and NTCP values as well as the interval on the TCP and NTCP curve could be presented. LET and variable-RBE maps could also provide complementary spatial information on biological effectiveness.

In addition, it would be valuable to evaluate the impact of probabilistic evaluation metrics on clinical decision making. This could involve presenting radiation oncologists with probabilistic evaluation metrics during treatment planning and systematically assessing how these metrics influence their decision making compared with conventional robustness evaluation. Preferably this would be done both for proton and photon planning, as most clinical experience and outcome data are derived from photon therapy.

This study included a relatively small number of radiation oncologists, medical physics experts and national centers. Future work could expand to an international setting with a larger and more diverse group of radiation oncologists and medical physics experts to further validate the findings. Such efforts are particularly relevant in light of ongoing ICRU work to revisit dose prescription, reporting and recording for advanced optimization strategies [38]. In addition, radiation oncologist interviews could be extended to photon planning to further assess the applicability of probabilistic evaluation metrics across treatment modalities.
Because the probabilistic outcomes depend on the assumed uncertainty distributions, the resulting probabilistic clinical goals should be interpreted within the context of the uncertainty model used in this study rather than as universal thresholds.

In conclusion, the results of this study indicate that probabilistic evaluation metrics are considered clinically usable and an improvement over current methods by radiation oncologists. There is a need for further consensus and clinical validation, both for protons and photons, before widespread implementation in clinical practice.